\documentclass[sigconf, screen]{acmart}
\AtBeginDocument{%
  \providecommand\BibTeX{{%
    \normalfont B\kern-0.5em{\scshape i\kern-0.25em b}\kern-0.8em\TeX}}}

\copyrightyear{2023}
\acmYear{2023}
\setcopyright{rightsretained}\acmConference[CHI '23]{CHI Conference on Human Factors in Computing Systems}{April 23-28, 2023}{Hamburg, Germany}
\acmBooktitle{CHI Conference on Human Factors in Computing Systems (CHI '23), April 23-28, 2022, Hamburg, Germany}
\acmConference[CHI 2023]{ACM CHI Conference on Human Factors in Computing Systems}{April 23--April 28}{Hamburg, Germany.}
\acmPrice{}
\acmDOI{}
\acmISBN{}

\usepackage{graphicx}
\usepackage[utf8]{inputenc}
\usepackage{amsmath}
\usepackage{units}
\usepackage{xcolor}
\usepackage{booktabs}
\usepackage{textcomp}
\usepackage{multirow}

\usepackage{microtype}        % Improved Tracking and Kerning
\usepackage{ccicons}          % Cite your images correctly!

\author{Kashyap Todi}
\email{kashyap.todi@gmail.com}
\affiliation{%
\institution{Meta Reality Labs}
\country{USA}
}

\author{Tanya R. Jonker}
\email{tanya.jonker@meta.com}
\affiliation{%
\institution{Meta Reality Labs}
\country{USA}
}

\begin{document}

\title{A Framework for Computational Design and Adaptation of Extended Reality User Interfaces}


% \renewcommand{\shortauthors}{x et al.}

\begin{abstract}
To facilitate high quality interaction during the regular use of computing systems, it is essential that the user interface (UI) deliver content and components in an appropriate manner.
Although extended reality (XR) is emerging as a new computing platform, we still have a limited understanding of how best to design and present interactive content to users in such immersive environments. 
Adaptive UIs offer a promising approach for optimal presentation in XR as the user's environment, tasks, capabilities, and preferences vary under changing context.
In this position paper, we present a design framework for adapting various characteristics of content presented in XR. 
We frame these as five considerations that need to be taken into account for adaptive XR UIs: \emph{What?}, \emph{How Much?}, \emph{Where?}, \emph{How?}, and \emph{When?}. 
With this framework, we review literature on UI design and adaptation to reflect on approaches that have been adopted or developed in the past towards identifying current gaps and challenges, and opportunities for applying such approaches in XR.
Using our framework, future work could identify and develop novel computational approaches for achieving successful adaptive user interfaces in such immersive environments. 
 % work can inform and inspire novel computational interaction research in this domain.

\end{abstract}

\begin{teaserfigure}
\centering
  \includegraphics[width=\textwidth]{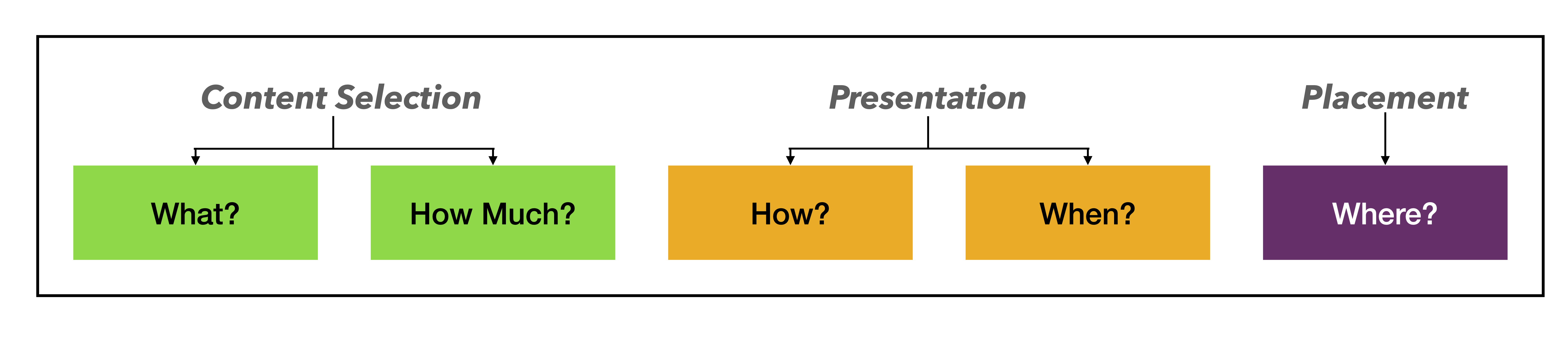}
  \caption{We propose a framework for design and adaptation of XR UIs. The framework addresses key questions related to content selection, presentation, and placement. By addressing these aspects, future XR systems can systematically determine content and layout of UIs that are optimal for varying contexts of usage.}
  \Description{A diagram illustrating our framework for design and adaptation of XR UIs. The framework addresses key questions related to content selection, presentation, and placement. By addressing these aspects, future XR systems can systematically determine content and layout of UIs that are optimal for varying contexts of usage.}
  \label{fig:teaser}
\end{teaserfigure}

%   <concept>
%       <concept_id>10003120.10003121.10003129</concept_id>
%       <concept_desc>Human-centered computing~Interactive systems and tools</concept_desc>
%       <concept_significance>500</concept_significance>
%       </concept>
%  </ccs2012>

% \ccsdesc[500]{Human-centered computing~Interactive systems and tools}

% \keywords{\plainkeywords}

\maketitle

\section{Introduction}

% XR combines real and virtual - new challenges for creating UIs 
% World-locked and head-locked UIs 
% Large design space - several possibilities 
% Adaptive-first approach - under changing context, what is a good UI also changes

% Prior work has looked into various aspects related to design and adaptation of UIs.
% Several computational approaches for achieving promising design solutions
% This has been done for varying domains -- from traditional 2D UIs to cross-device and 3D UIs.

% This paper motivated by a need to hollisticly understand adaptive behavior of UIs required for usable and performant XR interactions
% Develop a framework to characterize UI design and adaptation along five dimensions (or questions): what.... 
% Review prior literature accordingly -- based on properties they adapt and features that drive these adaptations.
% Summarize some computational approaches that have been applied towards optimal design and adaptation.
% Findings can inform future research directions and priorities 

Extended reality (XR) is a growing area of post-desktop computing, spanning virtual reality (VR), augmented reality (AR), and mixed reality (MR).
It offers the promise of a new computing environment for the future, where users are able to seamlessly interact with both the real and the virtual world. To realize this promise, XR requires a user interface (UI) that overlays virtual content onto or embeds it within the real, physical environment without costs to users' attention, effort, and safety, among other things. 
Prototypes of possible UIs include approaches that either take advantage of semantics shared between real and virtual content to associate UIs to the environment \cite{cheng2021semantic} or operate independently by assuming virtual content is anchored to the user \cite{damian2015augmenting}. 
Further, representation and modality are additional key aspects for rendering UIs in XR: information can be presented via non-visual modalities, such as audio and haptics; the size or level of detail can be modulated; and the same information can be presented visually using various two- or three-dimensional representations. 
As such, there is an immense design space for XR UIs. 

User's environment, task, and capabilities, often referred to collectively as \textit{context} \cite{lieberman2000context}, have large variance during continual use of XR systems and influence what a `good UI' might be. 
% Taking this into account, research has also identified the need for an adaptive-first approach to developing XR UIs.
Intermixing of real and virtual introduces substantial risk of distraction, additional attention costs, and even safety risks. 
As such, visual `real estate' for virtual content is (1) reduced and (2) constantly changing. 
Furthermore, as users' tasks and contexts continually change, their interest in and need for virtual interaction also constantly vary.
In contrast to typical computing platforms, where adaptive behavior often plays a small (and neglectable) role, it is believed that adaptive UIs might be the default for future XR environments \cite{adaptiveXRfuture, jonker2020role}.

Prior work has looked into various aspects related to design and adaptation of UIs, as applied to various domains ranging from traditional 2D menus and graphical UIs (e.g. \cite{gobert2019sam, lok2001survey, li2018layoutgan}) to cross-device \cite{park2018adam} and immersive 3D UIs \cite{lindlbauer2019context, grubert2017pervasive}.
Correspondingly, there have been several computational approaches, such as constraint solving \cite{zeidler2012constraint}, optimization \cite{o2014learning}, and machine learning \cite{li2018layoutgan}, developed and applied for achieving promising design solutions. 
While existing literature can provide valuable insights on approaches that can be adopted towards achieving certain adaptive behavior for future XR systems, current understanding of what aspects should be addressed when adapting such UIs is limited and fragmented.

This position paper is motivated by the need to holistically understand adaptive behavior of UIs that will be required for achieving highly usable and performant interactions in future XR environments. 
To this end, we introduce a novel framework that characterizes UI design and adaptation along five dimensions (or questions): \textit{what?}, \textit{how much?}, \textit{how?}, \textit{where?}, and \textit{when?}.
By answering these questions, adaptive XR systems could dynamically select appropriate content, presentation, and placement of UIs. 
We review prior literature on computational design and adaptation approaches across a range of application domains to understand what properties of the UI they address and features that drive adaptations in the UI.
We also summarize some of the technical approaches that have been developed and applied towards achieving optimal design and adaptation.
% For each of these properties, we discuss why they are essential for future XR systems and explore prior literature that has explored similar questions.
Our review highlights the need for further studying various UI adaptations directly in the context of extended reality, with varying tasks, environments, and capabilities. 
Our findings can inform future research directions in the area of adaptive XR UIs, driven by computational methods, to improve usability and user experience. 

\subsection* {Overview: Adaptive XR UI Framework}\label{sec:framework}

Context, including user's task, situational capabilities, and the environment, plays a key role in determining the right UI in XR environments \cite{lieberman2000context, lindlbauer2019context, jonker2020role} .
We develop a framework for designing and adapting XR UIs based on varying context.
In our framework, we consider different aspects that influence the final UI available to the user (\autoref{fig:teaser}).
This includes: (1) \emph{content selection} (\autoref{sec:selection}), which addresses \emph{what} content should be made available and \emph{how much} of it; (2) \emph{presentation} (\autoref{sec:presentation}), which refers to \emph{how} and \emph{when} content and components are presented; and (3) \emph{placement} (\autoref{sec:placement}), which determines \emph{where} UIs are positioned in the 3D environment.
% Further, we discuss computational approaches (\autoref{sec:computational}) that can be adopted by XR systems to achieve adaptive behavior.

\section{Content Selection: \textit{What?} and \textit{How Much?}}\label{sec:selection}

Selecting the \textit{right content} (`what?') to present to the user is necessary so as to ensure that users have important task-relevant elements readily available, while minimizing task-irrelevant distractions.
Further, the \textit{right amount}, or level of detail, of content (`how much?') should be modulated to ensure that users can interact at the appropriate granularity.

\textbf{What}: Prior works have explored several factors that can influence \textit{what?} should be presented to users in adaptive UIs, including: interaction history, user preferences, user capabilities, task, environment, aesthetics, and device capabilities.
Interaction history, containing prior usage of a system, has been used in traditional desktop-based environments to identify and select important items for adaptation based on features such as frequency, recency, and user interests \cite{gobert2019sam, bylinkskii2017learning, todi2018familiar, todi2021adaptive}.
Other works have investigated the role of user preferences \cite{park2018adam, schrier2008adaptive} and capabilities \cite{sarcar2018ability} to determine and adapt content selection to individual users.
Further, task \cite{julier20000filtering} and environment \cite{belo2022auit} are also deemed to be key factors that can inform what content should be presented to the user: for example, some applications may be used more consistently for certain tasks or some components have higher priority in certain environments.
Aesthetics can also influence content selection \cite{balinsky2009aesthetic}. 
In traditional 2D interfaces, aesthetics relates exclusively to how a selection of UI components can compose a harmonic and aesthetically pleasing UI; for XR, however, the harmonization of the real world and virtual elements can also influence content selection.
Finally, device capabilities can inform suitable content; this has been used, for example, to select content for presentation \cite{schrier2008adaptive, nichols2006uniform} or appropriate distribution across devices \cite{park2018adam, grubert2015multifi}.

\textbf{How Much}:
In addition to identifying what content should be selected, prior works have also investigated how much content is appropriate. `How much' content can be defined by properties such as number of items, level of detail, and information density.
For example, cognitive load \cite{sweller1988cognitive} is a key factor that influences usability and is closely related to how much content is presented to the user.
As such, prior work has used this as a measure to determine the appropriate granularity \cite{lindlbauer2019context}.
Task and environment are also key factors that have been used to determine how much content or UI should be presented \cite{schrier2008adaptive, diverdi2004detail, rao1994tablelens, grubert2017pervasive, julier20000filtering, oulasvirta2017combinatorial}.
As the context changes, the utility of various content changes too, which can be used to determine the amount of information presented.
Similarly, some prior works have considered user abilities and capabilities to drive adaptations or design variations \cite{gajos2007capibilities, julier20000filtering, sarcar2018ability}, and
others have looked at adapting information density, level of detail, or granularity based on varying device capabilities \cite{gajos2004supple, nichols2006uniform, diverdi2004detail, drexel2004bridging}.

In general, content selection is key to ensuring that desired content is readily available to users, while distracting and irrelevant content is minimized. 
Inappropriate selection can lead to increased cognitive load (e.g. due to excessive content or inappropriate granularity), additional interaction steps to retrieve unavailable content, and decreased performance when interacting with the UI.
As a first step of the adaptation process, we believe that future XR systems should first apply computational methods to optimally select content at the right level of detail.

\section{Presentation: \textit{How?} and \textit{When?}}\label{sec:presentation}

Content and UIs can often be presented to users via multiple representations.
For example, information can be presented textually or graphically, incoming notifications can be pushed using audio or haptics, or a continuous value selection component can be presented using a slider or a dial. 
Selecting the appropriate representation is crucial to ensure usability. 
Further, suitable timing for presenting content or adapting the presented content is important to ensure users are not surprised or confused by changes and continuity is maintained.    

\textbf{How}: 
Selecting the appropriate modality and representation for content or components can improve how users perceive and interact with the UI.
As such, this aspect has been studied by prior works in the context of 2D and 3D UIs.
One approach for selecting a representation is to consider other UIs -- from different applications with similar features or different devices with similar applications -- that are also being used: ensuring consistency or compatibility across UIs can make it easier for users to learn and understand the system \cite{nichols2006uniform, gajos2004supple}.
To improve aesthetics, color harmony across components has been studied in the context of UIs \cite{todi2016sketchplore, cohenor2006harmony}. 
For XR systems, in addition to harmony across virtual components, an additional consideration of harmonizing with the surrounding environment would be beneficial as the level of integration between virtual elements and the real world influences decision making and reaction time \cite{flittner2020predicting}.
Task and environment have also been used by prior works for determining appropriate representation of UIs \cite{bell2001view, nichols2006huddle}.
Finally, varying device and user capabilities have also influenced how content should be represented \cite{beshers1989scope, gajos2004supple, gajos2007capibilities, nichols2006uniform, nichols2006huddle}.

\textbf{When}:
Timing of content presentation or adaptation is another important factor that can determine how distracting, confusing, and usable the system might be; an inappropriately timed adaptation can at best surprise the user and at worst prevent them from completing their tasks.
Unlike other factors, to date, there is only a limited understanding of when changes should be triggered in adaptive systems.
Some prior works have used heuristics such as changes in the environment, task, or perceived cognitive load to determine appropriate timing \cite{lindlbauer2019context, madsen2016temporal, grubert2017pervasive}.
A more promising approach is to model and predict performance improvements that would be achieved if the system triggered an adaptation. 
This has been studied in the context of 2D menu-based UIs \cite{todi2021adaptive} and 3D XR UIs \cite{belo2022auit, yu2022timing} as a principled approach to adapt the system.

We hypothesize that the representation and timing aspects when presenting UIs to users in XR environments will largely influence how acceptable and usable these systems will be during all-day usage under varying contexts.

\section{Placement: \textit{Where?}}\label{sec:placement}
The last question or consideration for designing and adapting UIs in XR environments is their placement: where should content and UI components be placed such that users can interact with them with minimal effort?
As such, placement affects key usability aspects such as discoverability, reachability, exertion, and performance.
% Yet it is often challenging to optimally position elements when designing graphical layouts \cite{sketchplore}.
% The \emph{where?} in our framework relates to this positioning aspect: where should components be placed so that users can interact with them with minimal effort?
Determining placement is an especially hard problem as special attention needs to be paid to relative placement of components to other components, in addition to each element's absolute position.
This includes aspects such as sequential or logical ordering, reading order, and semantic relationships between components.
% XR introduces the additional challenge of placing UIs in relation to objects and features in the real environment. 

A wide range of factors have been used by prior research to determine where UIs and content should be placed, in the context of 2D, 3D, and cross-device UIs.
Constraint-based layouts, where relationship between components are defined using constraints, have been one common approach to determine final placement of elements \cite{hosobe2001constraint, schrier2008adaptive, badros2001cassowary, gal2014flare}. 
Similarly, abstract UI specifications have been used to generate final UI placement for varying context \cite{gajos2004supple, beshers1989scope, nichols2006uniform}.
This enables systems to position components while maintaining consistency when features such as screen size and aspect ratio vary.
Grid and geometric based approaches have also been widely used \cite{jacobs2003grid, dayama2020grids, hosobe2001constraint, balinsky2009aesthetic}: here, structural properties are captured using mathematical formulation to ensure properties such as alignment and rectangularity.
Occlusion avoidance and ensuring viewability also influence placement decisions and have been used to generate or adapt UIs \cite{cheng2021semantic, hosobe2001constraint, azuma2003label, bell2001view, kohei2008viewability, belo2022auit}.
Some prior works have made placement decision based on other relative placement in other UIs to ensure consistency and similarity \cite{todi2018familiar, kumar2011bricolage, nichols2006uniform, li2018layoutgan}. 
Appropriate placement in UIs also highly correlates to task performance and perceived aesthetics. 
As such, interactive systems have studied UI placement to optimize these qualities \cite{todi2016sketchplore, bailly2013menuoptimizer, lok2004balance, balinsky2009aesthetic, o2014learning, belo2021xrgonomics}.  
Finally, user task and environment has also been utilized as a factor that determines where components are placed \cite{lindlbauer2019context, cheng2021semantic, madsen2016temporal, fender2018optispace, fender2017heatspace, nuernberger2016snap, bell2001view, belo2022auit, nichols2006huddle, bordes2011adaptive}.

In XR, content and components do not occupy a dedicated screen or canvas; instead, they are overlaid on the real world and can span the entire 360\textdegree~ environment around the user.
Further, virtual components can not only have semantic relationships with each other, but with real world objects too.
As such, typical approaches such as grid- and constraint- based layouts can not be directly applied.
Recently, some research has explored how aspects such as task \cite{lindlbauer2019context} and semantic relationships \cite{cheng2021semantic} can be used to adaptive determine where elements should be placed in XR environments.
We suggest that future XR systems will need to carefully address placement issues by considering various aspects pertaining to both the user and the environment, such as cognitive load, ergonomics, task performance, aesthetics, semantic consistency, user task and capabilities, and device constraints. 

% \section{Computational Approaches for Adaptation}\label{sec:computational}

% \begin{enumerate}
%     \item Heuristics
%     \item Statistical modeling
%     \item Constraint satisfaction
%     \item Optimization 
%     \item Machine Learning
% \end{enumerate}

\section {Discussion}

In this position paper, we have studied design and adaptation of UIs in the context of extended reality (XR) systems.
We introduce a framework consisting of five dimensions -- \textit{what?}, \textit{how much?}, \textit{how?}, \textit{where?}, and \textit{when?} -- to characterize various factors that can influence the final quality of a UI. 
By addressing each of these, future XR systems could systematically select appropriate content, manipulate their presentation, and place them to optimally support interactions.
We discuss existing literature on computational approaches for UI design and adaptations that touch upon these questions to understand the state-of-the-art and identity opportunities for future research.
By discussing prior work, we can identify gaps, challenges, and opportunities for novel contributions towards making adaptive-first XR a reality. 
Firstly, it is evident that we need to carefully consider the blending of the real environment and virtual UIs when making decisions related to content selection and presentation; this includes aspects related to both performance and aesthetics.
Second, as situational capabilities vary with changing contexts, this is another key aspect that should be studied when determine adaptive system behavior.
Finally, inherently adaptive systems will need to pay more attention to timing of changes and content representations that are suitable to the user's context.

As the computing paradigm shifts from strict separation of the real world from virtual interactive interfaces to a blending of these environments in extended reality, we face new and amplified challenges related to developing design and development of UIs. 
While there is extensive prior research on computational design and adaptation of UIs that can be applied to XR scenarios, our understanding so far is fragmented and targeted towards specific challenges. 
We hope that our design framework can provide new insights that lead to a structured approach for adaptive XR systems of the future that tackle key considerations holistically.

\bibliographystyle{ACM-Reference-Format}
\bibliography{References.bib}

%\appendix

\end{document}